\documentclass[12pt,preprint]{aastex}

\slugcomment{Accepted to Astrophysical Journal}

\shorttitle{24 \micron Radius of HD 209458 b}
\shortauthors{L.\ J.\ Richardson et al.}

\begin{document}


\title{A {\it Spitzer\footnote{This work is based on observations made with the Spitzer Space Telescope, which is operated by the Jet Propulsion Laboratory, California Institute of Technology under a contract with NASA. Support for this work was provided by NASA.}}\, Infrared Radius for the Transiting Extrasolar Planet HD\,209458\,b}
%\title{The 24 Micron Radius of the Transiting Extrasolar Planet HD\,209458\,b}
%\title{24-Micron Transit Observations of the Extrasolar Planet HD\,209458\,b}


\author{L.~Jeremy Richardson}%\altaffilmark{2} }

\affil{Exoplanets and Stellar Astrophysics Laboratory, NASA's Goddard 
Space Flight Center, Mail Code 667, Greenbelt, MD 20771}
\email{richardsonlj@milkyway.gsfc.nasa.gov}

\author{Joseph Harrington}
\affil{Center for Radiophysics and Space Research, Cornell Universtiy, 
326 Space Sciences Bldg., Ithaca, NY  14853-6801}
\email{jh@oobleck.astro.cornell.edu}

\author{Sara Seager}
\affil{Department of Terrestrial Magnetism, Carnegie Institution of 
Washington, 5241 Broad Branch Rd NW, Washington, DC  20015}
\email{seager@dtm.ciw.edu}

\author{Drake Deming}
\affil{Planetary Systems Laboratory, NASA's Goddard Space Flight Center, 
Mail Code 693, Greenbelt, MD 20771}
\email{ddeming@pop600.gsfc.nasa.gov}

%\altaffiltext{2}{NRC Postdoctoral Research Fellow}

\begin{abstract}
We have measured the infrared transit of the extrasolar planet
HD\,209458\,b using the Spitzer Space Telescope.  We observed two
primary eclipse events (one partial and one complete transit) using
the 24~\micron\ array of the Multiband Imaging Photometer for Spitzer
(MIPS).  We analyzed a total of 2392 individual images (10-second
integrations) of the planetary system, recorded before, during, and
after transit.  We perform optimal photometry on the images and use
the local zodiacal light as a short-term flux reference.  At this long
wavelength, the transit curve has a simple box-like shape, allowing robust solutions for the stellar and planetary radii independent of stellar limb darkening, which is negligible at 24~\micron.  We derive a stellar radius of
R$_*$ = 1.06 $\pm$ 0.07~R$_\sun$, a planetary radius of
R$_p$ = 1.26 $\pm$ 0.08~R$_{\mbox{\scriptsize J}}$, and a stellar mass of 1.17~M$_\sun$.  Within the errors, our results agree with the measurements at visible wavelengths.  The 24~\micron\ radius of the planet therefore does not differ significantly compared to the visible result.  We point out the potential for deriving extrasolar transiting planet radii to high accuracy using transit photometry at slightly shorter IR wavelengths where greater photometric precision is possible.

\end{abstract}


%\keywords{globular clusters: general ---
%globular clusters: individual(\objectname{NGC 6397},
%\object{NGC 6624}, \objectname[M 15]{NGC 7078},
%\object[Cl 1938-341]{Terzan 8})}


\section{Introduction}

The transit of an extrasolar planet across its star allows us to
measure the radius of the planet \citep{charbonneau00, henry00, brown01a}.  Of the ten known transiting planets, HD\,209458\,b, with a radius (at visible wavelengths) of 1.320 $\pm$ 0.025~R$_{\mbox{\scriptsize J}}$ \citep{knutson06}, is inflated compared to the other known transiting planets and thus has a lower bulk density.  One explanation for the anomalous radius was inflation by the dissipation of tidal stress within the planet \citep{bodenheimer01}.  However, the timing
of the secondary eclipse as observed by the Spitzer Space Telescope
\citep{deming05b}, as well as improved radial velocity observations
\citep{laughlin05}, have ruled out a non-zero orbital eccentricity of
the magnitude ($\sim$0.03) needed by the tidal dissipation theory.
%Other explanations involve the fact that the observed radius is a transit radius whose value is sensitive to the opacity of the upper atmosphere \citep{burrows03}.
\citet{showman02} suggest that kinetic energy produced by atmospheric circulation and deposited in the planet's interior could account for the missing energy source and increase the planetary radius.
Another popular proposed explanation is the possibility of obliquity tides \citep{winn_holman05}, in which a non-zero obliquity (made possible by a spin-orbit resonance) could drive the tidal dissipation and provide the necessary energy to give the planet an inflated radius.

Our Spitzer program to measure the 24~\micron\ flux of HD\,209458\,b
includes observations during transit (i.e., primary eclipse), revealing the infrared (IR) radius of the planet, which is reported in this paper.
Section~\ref{sec:motiv} further elaborates on why an IR
radius measurement is of interest.  Section~\ref{sec:obs} describes
the observations; section~\ref{sec:analysis} explains the photometric
data analysis and radius fit.  Section~\ref{sec:disc} concludes with
results and discussion.

\section{Motivation for Infrared Radius Measurements}
\label{sec:motiv}

The reduced stellar flux at mid-IR wavelengths implies that transit
photometry in this region is unable to achieve the high
photometric precision obtained at visible wavelengths \citep{brown01a}.  However, stellar limb darkening weakens with increasing wavelength due to the increasing H$^{-}$ free-free opacity \citep{vernazza76}. Thus, the fitting
of transit curves to mid-IR data is simple and robust, and gives results
independent of limb-darkening parameterizations.  A mid-IR radius
measurement is also of intrinsic interest for understanding the
planet.  
Observations of the planet during transit suggest that clouds and scattering layers could potentially extend to great heights.  The low sodium abundance reported by \citet{charbonneau02} and the upper limit on CO reported by \citet{deming05a} both support this physical picture of the planet's atmosphere.  Furthermore, the escaping atmosphere observed by \citet{vidal-madjar03} suggests that other processes may be at work in planet's atmosphere.
The scale height of the atmosphere is $\sim$450~km, and if the clouds extend to even a few scale heights, this would represent $\sim$0.01~R$_{\mbox{\scriptsize J}}$.  While our observations are not able to reach that level of precision in the planetary radius, it is nonetheless valuable to search for unexpectedly large variations in the planetary radius as a function of wavelength.  %Scattering opacity due to small particles will be greatly reduced at long wavelengths, implying that the radius could be measurably smaller.  
A major goal of our observational program was to test this scenario by measuring the 24~\micron\ radius of HD\,209458\,b.

\section{Observations}
\label{sec:obs}

We observed two transit events using the Spitzer Space Telescope \citep{werner04}: a half eclipse event (ingress only) on 2004 December 5 and a full eclipse event on 2005 June 27.  We used the MIPS \citep{rieke04} 24~\micron\ array, which is a Si:As detector with 128$\times$128 pixels and an image scale of 2.55 arcsec pixel$^{-1}$.
We obtained two series of 10-second exposures using the standard MIPS raster pattern, which places the star at 14 different positions on
the array.  This produced 864 images during the half eclipse event and
1728 images during the full eclipse event.

%\section{Photometry and Light Curve Fitting}
\section{Analysis}
\label{sec:analysis}

\subsection{Photometry}

We first reject obviously bad images, including those with strong
`jailbar' features\footnote{See the MIPS Data Handbook, available at {\tt http://ssc.spitzer.caltech.edu/mips/dh/}}, 
as well as the initial image of each cycle due to
a prominent `first frame' effect.  This leaves 780 images for the half
eclipse event, and 1612 images for the full eclipse event.  We analyze
each eclipse event separately.  For each of the 14 raster positions,
we perform the following steps:
\begin{enumerate}
\item {\em Median filter the images to remove energetic particle events and hot pixels.}  The median image is computed from all the images (typically 60) at a given raster position, and the difference image is constructed by subtracting the median image from a given image.  We then employ a routine called SIGMA\_FILTER\footnote{See the IDL Astronomy Library at {\tt http://idlastro.gsfc.nasa.gov/contents.html}}, which computes the mean and standard deviation of the pixels within a box of specified width (excluding the center pixel).  If the center pixel is deviant by more than a specified number of standard deviations from its neighbors, it is replaced by the mean of the remaining pixels in the box.  We use a box width of 20 pixels and a sigma limit of 10, and iterate until no more pixels are changed.  This cleans most of the hot pixels evident in a given image.  Rejected from further consideration are any images in which a pixel within the defined aperture containing the star is changed.
\item {\em Calculate and subtract the total zodiacal background from all pixels in a given image.}
%except those in a small (3$\times$3 pixel) box surrounding the star.
The background level for each image is determined by constructing a histogram of all pixels in an image (with a bin size of 0.01 MJy/sr) and fitting a Gaussian to the histogram.  The center of the Gaussian fit is then the `average' background level for that image, and this constant level is subtracted from the each pixel in the image to create the background-subtracted image.
\item {\em Find the center of the star to a precision of 0.01~pixel by dithering the theoretical point spread function (PSF) over the individual images and finding the best fit.}  The theoretical PSF was obtained\footnote{See {\tt http://ssc.spitzer.caltech.edu/mips/psffits/} and the file {\tt mips\_24\_5000c.fits}.} for a 5000~K blackbody on the center of the 24~\micron\ array.  These files have been modeled using Tiny Tim \citep{krist93}.  We first resample the theoretical PSF to a scale 100 times finer for comparison to the real data.  %, and we next resample by another factor of 25 (for an overall scale factor of 100). 
The resampled PSF is dithered in both dimensions, and the `best-fit' PSF to the data (using linear least squares) determines the center of the star.
\item {\em Use the best-fit PSF to weight the pixels near the star before adding them.}  This is applied to the background-subtracted images and produces optimal photometry \citep[analogous to][]{horne86}.  The errors are derived by propagating the MIPS errors through the optimal error formula \citep[][see his Table~1, item 8]{horne86}.  The errors are dominated by statistical fluctuations in the zodiacal background.  The optimal photometry typically provides a S/N improvement of 50\% or more over the standard aperture photometry approach (where the pixels are summed with no weighting function).
\item {\em Normalize the optimal photometry to the total background level in the frame.} The total background level is simply the average of all the points (except those in a 3$\times$3 box surrounding the star) in a given image.  This gives the stellar intensity relative to the zodi and thus removes any remaining instrument response variations.
\end{enumerate}
Having completed these steps for each raster position, we recombine the data to obtain the entire time series for each eclipse event.  Time is converted to orbital phase using the most recent and most accurate orbital period and ephemeris \citep{knutson06}, and we account for the light travel time between the Sun and Spitzer.  The zodiacal background changes linearly by $\sim$1.5\% over the six-hour duration over which the full eclipse event is observed, and we remove this effect from both eclipses separately.  We estimated the magnitude of this change using the Spitzer Observations Planning Tool (SPOT), which employs the zodiacal dust model from \citet{kelsall98}.  Finally, the time series are normalized to an average of unity for the out-of-transit points.

\begin{figure}
\begin{center}
\includegraphics{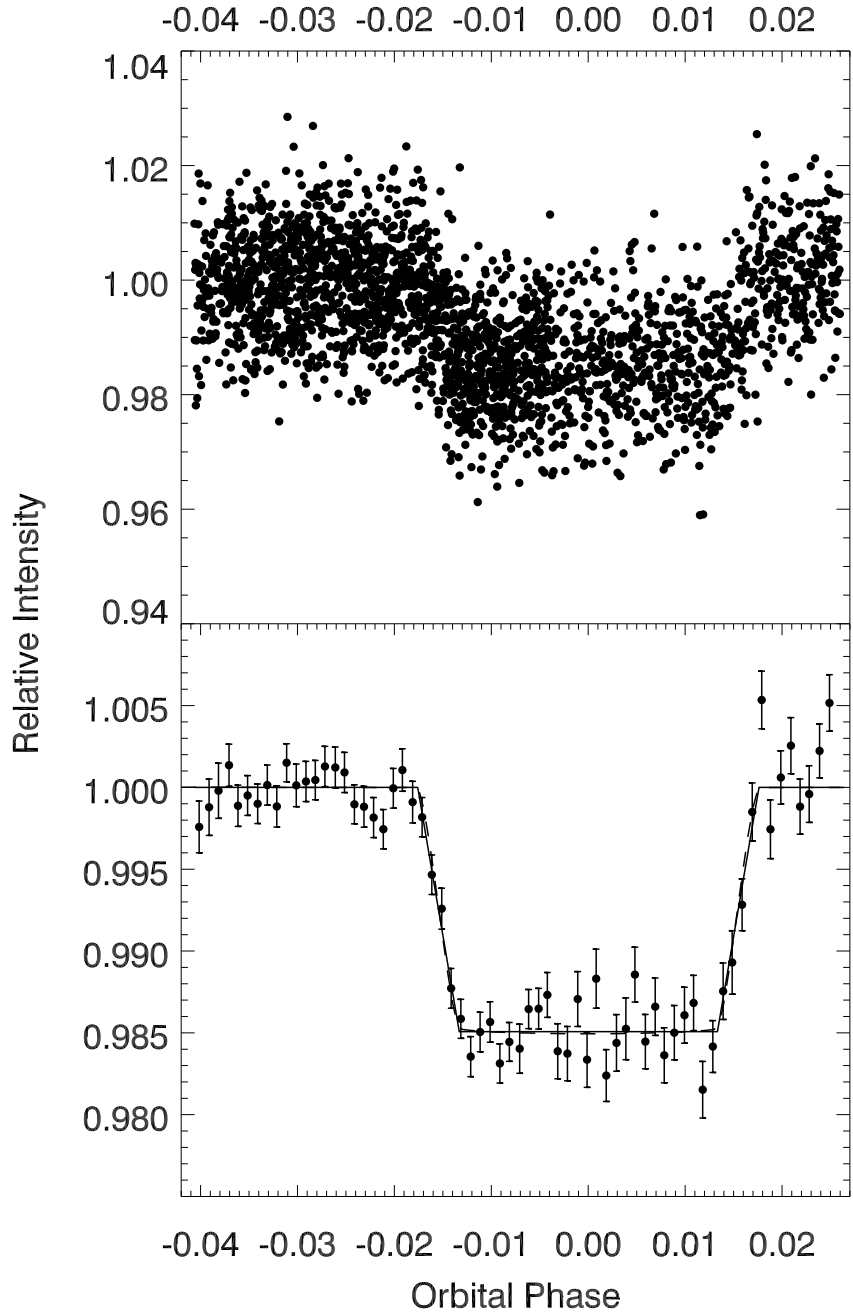}
\caption{
\emph{Upper Panel:} All 2392 measurements versus heliocentric phase.
\emph{Lower Panel:} Data averaged in phase (bin size = 0.001 in phase); also shown are the best fit straight-line curve (solid line) and the exact theoretical light curve (dashed line, difficult to see), calculated using the best-fit physical parameters derived from the straight-line curve.  Heliocentric phase was computed using the orbital period and ephemeris from \citet{knutson06}.  
\label{fig:data} }
\end{center}
\end{figure}

The calibrated, unbinned photometry is given in Table~\ref{tbl:phot} and is shown in Figure~\ref{fig:data} (upper panel).  The upper panel
shows the aggregate data for both events combined (2392 points), and it clearly reveals the eclipse. 
%The scatter in the points is closely consistent with the errors calculated by propagating the per-pixel errors from the Spitzer pipeline data analysis through our optimal photometry procedure.  
The lower panel shows the average in bins of
phase width 0.001; the box-like shape of the light curve due to the lack of stellar limb darkening is quite evident.  To perform the optimal photometry, we used two independent codes, derived from the same basic algorithm but constructed by individual researchers. We obtained virtually identical per-point results with both.

\subsection{Light Curve Fitting}

\begin{figure}
\begin{center}
\includegraphics{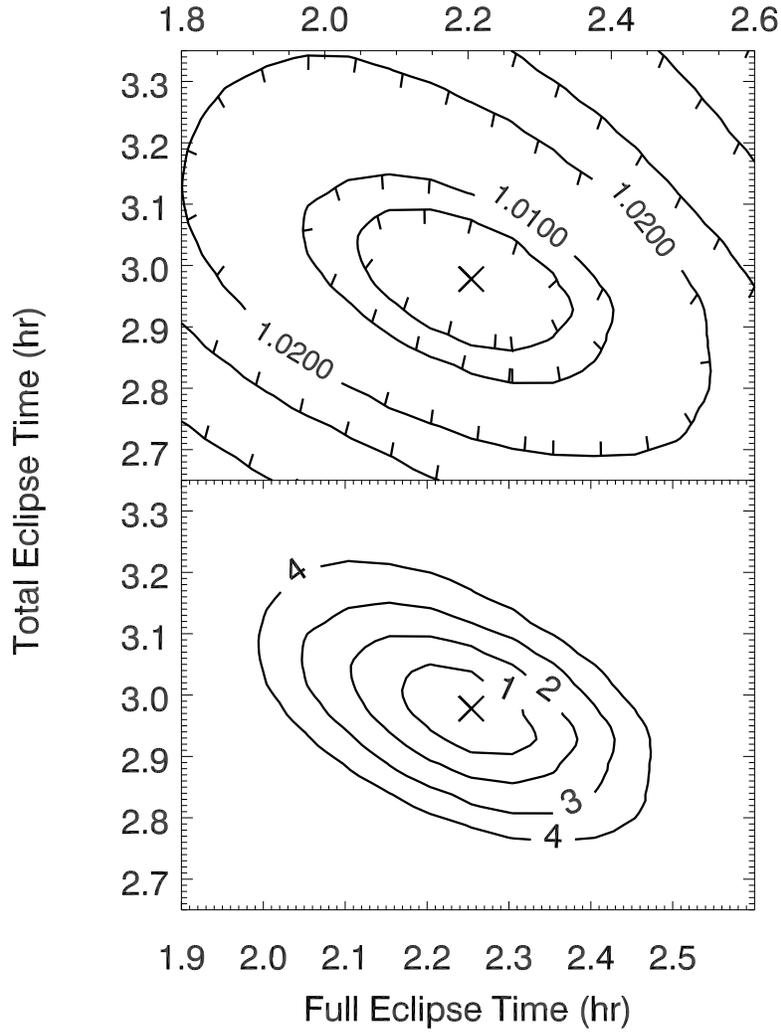}
\caption{
\emph{Upper Panel:} Contour plot of the reduced chi-squared ($\chi_\nu^2$) fit from Method~1, showing the total eclipse time vs.\ the full eclipse time at zero phase offset (best-fit value).  
\emph{Lower Panel:} Same result, but converted to confidence interval in standard deviations.  In both panels, the minimum $\chi_\nu^2$ (1.0051) is marked by an X.
\label{fig:cont} }
\end{center}
\end{figure}

We construct a family of simple, approximate light curves by connecting intensities at the contact times with straight line segments.  Four observable parameters uniquely describe a light curve in the absence of stellar limb darkening:  1) the duration of full eclipse (i.e., the time between second and third contact, $t_{\mbox{\scriptsize F}}$), 2) the total duration of the eclipse (i.e., the time between
first and fourth contact, $t_{\mbox{\scriptsize T}}$), 3) the eclipse depth, and 4) the observed time of center eclipse.  We derive best-fit values for these four observables by minimizing the reduced chi-squared ($\chi_\nu^2$) of the fits to each simple transit curve generated.  We refer to this `trial-and-error' technique as `Method~1.'  We have chosen a large enough range of parameters and small enough grid spacing to avoid finding only a local minimum.  Figure~\ref{fig:cont} shows $\chi_\nu^2$ as a function of $t_{\mbox{\scriptsize F}}$ and $t_{\mbox{\scriptsize T}}$ for zero phase offset (the best-fit value); the upper panel shows the contour plot for the $\chi_\nu^2$, and the lower panel casts the result in terms of confidence intervals.  (The contours and confidence intervals are correctly calculated by projecting the $\chi_\nu^2$ into the plane of interest; see \citet[][Section~15.6 and Figure~15.6.4]{press92}.)  
The best-fit observables ($t_{\mbox{\scriptsize T}}$, $t_{\mbox{\scriptsize F}}$, eclipse depth, and time offset) are listed in Table~\ref{tbl:obs} (column marked `Method~1').

In order to verify the results from Method~1 and to ensure that we did not hit a local minimum, we employed a second method of finding the best-fit observables to the data, the MPFIT package\footnote{{\tt http://cow.physics.wisc.edu/$\sim$craigm/idl/idl.html}}, which performs a Levenberg-Marquardt least-squares fit.  Combined with our own function that computes a theoretical eclipse (as described above), this method independently calculates the four observables, and these results are also shown in Table~\ref{tbl:obs} (column marked `Method~2').  We note that the two results agree closely, and we adopt the results from MPFIT (Method~2) as our formal results.  Although the resulting minimum $\chi_\nu^2$ is slightly larger, we nonetheless adopt Method~2 because it is computationally more efficient for calculating the errors on the individual parameters, as discussed below.

Using the analytic formulation of \citet{seager_mallen-ornelas03}, we derive the physical parameters of the system from the observables ($t_{\mbox{\scriptsize T}}$, $t_{\mbox{\scriptsize F}}$, and eclipse depth, obtained from the best-fit simple light curve, and the known orbital period \citep{knutson06}).  The impact parameter $b$, the ratio of the semi-major axis to the stellar radius, $a/R_*$, the orbital inclination $i$, and the stellar density $\rho_*$ are derived immediately from these observables \citep[][see their Equations 7, 17--19]{seager_mallen-ornelas03}\footnote{Note that their Equation 19, derived from Equation 9, is missing a factor of $\frac{4}{3}\pi$.}.

Determining the stellar and planetary radii from these parameters requires an assumption of the stellar mass \citep{brown01a}.
We assumed stellar masses between 0.9 and 1.30~M$_\sun$ (shown in Figure~\ref{fig:mstar}), covering a region surrounding the reported stellar mass of 1.146~M$_\sun$ \citep{brown01a}.  %Results for three of those assumed values are given in Table~\ref{tbl:params}, corresponding to those tabulated by \citet[][their Table~4]{wittenmyer05}.  
Interestingly, by using the analytic formulation, we note that the orbital inclination is determined by the transit times and the \emph{ratio} of the stellar radius to the orbital semi-major axis; this means that our assumption of the stellar mass, while determining the stellar radius, does not affect the orbital inclination.%, as seen in Table~\ref{tbl:params}.

We use a bootstrap Monte Carlo method \citep[][see Section~15.6]{press92} to determine realistic errors in the observables and physical parameters.
%We construct 1000 `synthetic' light curves using the best-fit observables and Gaussian noise with the same per-point errors as the real data.  In other words, we start with a straight-line-segment light curve with the best-fit observables, and add Gaussian noise with the same standard deviation as the real data; this produces a noisy light curve.  
%We start with the straight-line-segment fit (using Method 2), and subtract this fit from the real data (N=2392 points) to obtain the residuals.  
We randomly select N=2392 data points with replacement (meaning some points are duplicated) to create a `synthetic' data set.  We create 10000 such synthetic data sets, perform the same fitting procedure described above (again, Method 2, using MPFIT) to each one, and derive the physical parameters from the best-fit observables.  In this way, we derive a set of physical parameters (R$_*$, R$_p$, and $i$) for each of the 10000 synthetic data sets.  Using the routine HISTOGAUSS (from the IDL Astronomy Library\footnote{{\tt http://idlastro.gsfc.nasa.gov/contents.html}}), we fit a Gaussian to each one of the arrays of observables and physical parameters.  All parameters are normally distributed and symmetric, so that the width of each best-fit Gaussian represents the 1-$\sigma$ error in the associated parameter, and these are the uncertainties presented in Tables~\ref{tbl:obs} and~\ref{tbl:params}.  %The results for the four observables are plotted against each other individually in Figure~\ref{fig:cont}.

The bootstrap method was also performed on each of the eclipse events separately to determine the observed time of center eclipse, as shown in Table~\ref{tbl:ecltime}.  Both are consistent with zero offset in time from the predicted value.  For the half eclipse (event~1), we set the full eclipse time ($t_{\mbox{\scriptsize F}}$) to the value derived from fitting the aggregate data and hold it fixed, while minimizing the other three observables.  As expected, the uncertainty in the time of center for the half eclipse is much larger than that of the full eclipse.

Next, we check the radii and orbital inclination by removing the
approximation that the light curve is comprised of straight line
segments.  We have developed a routine to compute light curves numerically \citep{richardson05p2}, and we include the small effect due to predicted limb darkening at 24~\micron, derived from a Kurucz\footnote{Available from \url{http://kurucz.harvard.edu}.  We linearly interpolate the Kurucz parameters at 20 and 40~\micron\ to estimate the values at 24~\micron.}
model atmosphere for stellar parameters $T_e = 6000$~K, $\log{g} = 4.5$, and [Fe/H]$ = 0.0$.
We validated the code by verifying that we can reproduce the fits to the very precise HST optical data from \citet{brown01a}.  We adopted the derived  parameters from the best-fit simple curve and calculated an exact theoretical light curve.  The result is plotted as a dashed line in the lower panel of Figure~\ref{fig:data}, but it is hard to see since it is nearly identical to the simple curve.  The $\chi_\nu^2$ of the fits for each of the two curves to the data are nearly identical:  1.0060 for the best-fit simple curve from MPFIT, compared to 1.0061 for the theoretical light curve.  We therefore conclude that the limb darkening is negligible at 24~\micron\ (as expected) and that the simple light curve composed of straight line segments is an accurate method of deriving the physical parameters.

Finally, we checked our results by incorporating information from the transit at visible wavelengths \citep{brown01a}.  There the transit depth is 0.0164, compared to $0.0149 \pm 0.0003$ at 24~\micron, a significant difference.  The ratio of visible to IR transit depth can be used to determine the impact parameter, the minimum projected radius where at which the planet crosses the star, and thereby the orbital inclination.  That is, we are deriving the degree of limb darkening at the given projected stellar radius of closest approach and using the Kurucz model to determine the location of the chord.
We used the limb darkening tabulated by a Kurucz model atmosphere (same parameters used above), and the `small planet' approximation from \citet{mandel02}.  We calculated numerically the ratio of visible to IR transit depth as a function of impact parameter.  Comparing the observed ratio ($1.100 \pm 0.03$) to this relation gives an impact parameter of $0.58 \pm 0.07$,  and an orbit inclination of $86.6^\circ \pm 0.6^\circ$.  Within the errors, this agrees with the results at visible wavelengths \citep{brown01a,wittenmyer05} and is consistent with the $i=87.97^\circ \pm 0.85^\circ$ value we derive internally from our IR data.  This calculation serves as an independent check of our results and a direct comparison to the visible results.

\section{Results and Discussion}
\label{sec:disc}
\begin{figure}
\begin{center}
\includegraphics[scale=0.75]{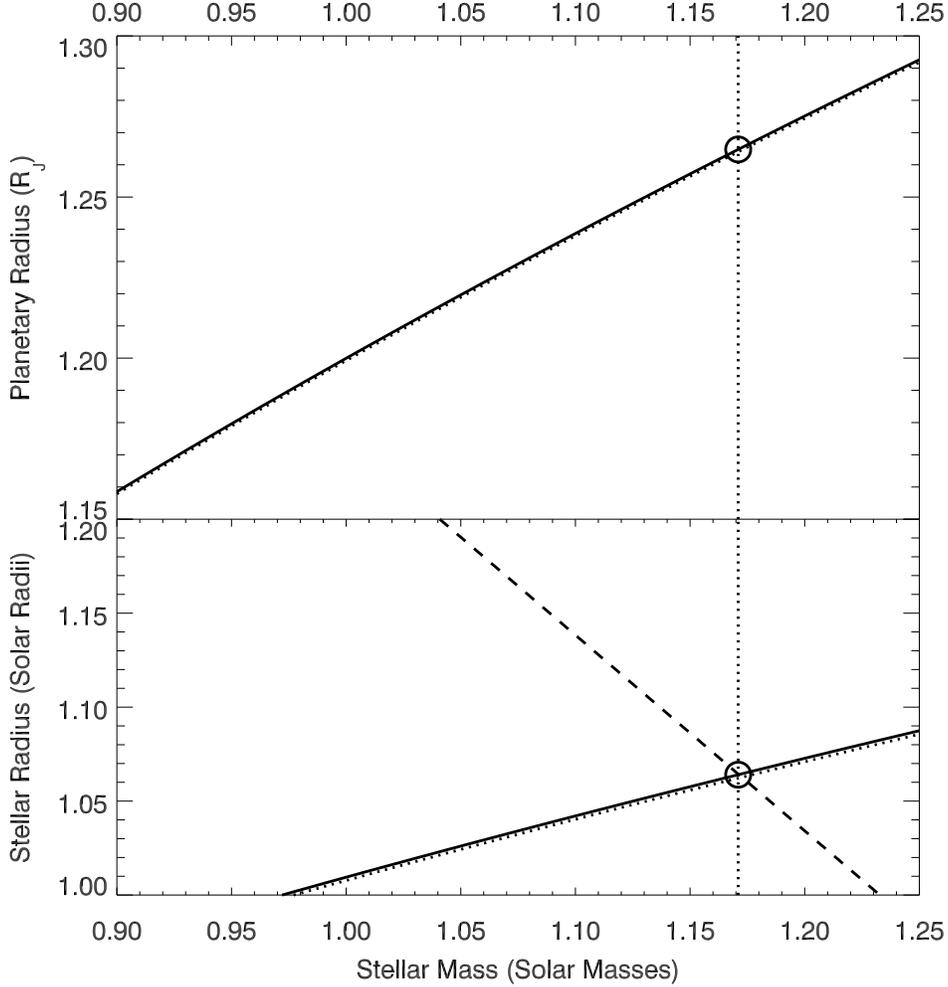}
\caption{The stellar density derived from the best-fit parameters can be transformed to stellar radius by assumption of a range of stellar masses.  The transit data therefore yield an empirical mass-radius relation.
\emph{Upper Panel:}  Planetary radius as a function of assumed stellar mass.  Solid line represents result from MPFIT (Method 2); dotted line represents the `trial-and-error' minimization technique (Method 1).
\emph{Lower Panel:} Stellar radius as a function of assumed stellar mass.  Methods~1 and 2~indicated as in upper panel.  Dashed line represents mass-radius relation from \citet{cody02}.
%dot-dash line is the mass-radius relation for a zero-age main sequence star \citep{claret89}. 
Intersection of this relation with the empirical curve to the assumed stellar masses allows a determination of the stellar mass (1.171~M$_\sun$), marked by the vertical dotted line.  This reveals the best fit stellar and planetary radii, R$_*$ = 1.06~R$_\sun$ and R$_p$ = 1.26~R$_{\mbox{\scriptsize J}}$.
\label{fig:mstar} }
\end{center}
\end{figure}

We have computed the stellar density directly from the observable quantities from the best-fit simple curve \citep{seager_mallen-ornelas03}.  Assuming a stellar mass allows us to calculate the stellar radius.  This empirical mass-radius relation is shown in Figure~\ref{fig:mstar}, where we have derived the radii for stellar masses from 0.9 to 1.3~M$_\sun$.
We break the degeneracy by intersecting the stellar radius curve with the mass-radius relation from \citet{cody02}, which is derived by fitting stellar models to a constant luminosity.  This is shown as the dashed line in Figure~\ref{fig:mstar}, following \citet[][their Figure 5]{wittenmyer05}.  On this basis, we derive the stellar mass to be M = 1.171~M$_\sun$, with R$_*$ = 1.06 $\pm$ 0.07~R$_\sun$ and R$_p$ = 1.26 $\pm$ 0.08~R$_{\mbox{\scriptsize J}}$.  Our result for the planetary radius agrees with the updated visible radius of R$_p$ = 1.320 $\pm$ 0.025~R$_{\mbox{\scriptsize J}}$ \citep{knutson06}.
%Our result for the planetary radius agrees with the results at visible wavelengths: R$_p$ = 1.347 $\pm$ 0.060~R$_J$ \citep{brown01a} and 1.35 $\pm$ 0.07~R$_J$ \citep{wittenmyer05}.
%Also shown is the mass-radius relation for a zero-age main sequence star from \citet{claret89}, which intersects the empirical relation at the same point.  Such exact agreement is probably fortuitous but suggests that the determination of the radii is not strongly dependent on the age of the star.

We are encouraged by the fact that our radius error is only four times larger than that obtained by \citet{knutson06}, in spite of the fact that our infrared photometric precision is an order of magnitude poorer than the HST visible photometry.
%We are encouraged by the fact that the photometric precision we obtain for the IR radius of the planet is similar to the current results at visible wavelengths \citep{wittenmyer05}, in spite of the fact that the visible photometric observations are more than an order of magnitude more precise than our 24~\micron\ photometry.
We attribute this to the character of the IR transit curve, where the lack of limb darkening produces a simple transit shape, from which radius information can be extracted with maximum efficiency.  We point out that photometry at other accessible Spitzer wavelengths such as 8 and 16~\micron\ would provide much higher photometric precision for bright transiting systems, because the stellar flux will be much higher, and the zodiacal background will not be a limiting factor.   Limb darkening remains sufficiently weak at these shorter IR wavelengths to maintain a relatively simple transit light curve shape.  Considering also that Spitzer's heliocentric orbit allows uninterrupted observations of complete transits, we suggest that IR transit photometry from Spitzer may be the optimal method for precise radius determination in bright transiting planet systems.

\acknowledgments
This work is based on observations made with the Spitzer Space
Telescope, which is operated by the Jet Propulsion Laboratory,
California Institute of Technology under a contract with NASA. Support
for this work was provided by NASA.  LJR acknowledges support as a NASA Postdoctoral Fellow (formerly NRC Research Associate) at NASA Goddard.
We thank the referee for insightful comments and suggestions that significantly improved the manuscript.

Facilities: \facility{Spitzer(MIPS)}

%%Facilities: \facility{Nickel}, \facility{HST(STIS)}, \facility{CXO(ASIS)}.




%\bibliographystyle{apj}
%\bibliography{refs,richardson}

\begin{thebibliography}{24}
\expandafter\ifx\csname natexlab\endcsname\relax\def\natexlab#1{#1}\fi

\bibitem[{{Bodenheimer} {et~al.}(2001){Bodenheimer}, {Lin}, \&
  {Mardling}}]{bodenheimer01}
{Bodenheimer}, P., {Lin}, D.~N.~C., \& {Mardling}, R.~A. 2001, \apj, 548, 466

\bibitem[{{Brown} {et~al.}(2001){Brown}, {Charbonneau}, {Gilliland}, {Noyes},
  \& {Burrows}}]{brown01a}
{Brown}, T.~M., {Charbonneau}, D., {Gilliland}, R.~L., {Noyes}, R.~W., \&
  {Burrows}, A. 2001, \apj, 552, 699

\bibitem[{{Charbonneau} {et~al.}(2000){Charbonneau}, {Brown}, {Latham}, \&
  {Mayor}}]{charbonneau00}
{Charbonneau}, D., {Brown}, T.~M., {Latham}, D.~W., \& {Mayor}, M. 2000, \apjl,
  529, L45

\bibitem[{{Charbonneau} {et~al.}(2002){Charbonneau}, {Brown}, {Noyes}, \&
  {Gilliland}}]{charbonneau02}
{Charbonneau}, D., {Brown}, T.~M., {Noyes}, R.~W., \& {Gilliland}, R.~L. 2002,
  \apj, 568, 377

\bibitem[{{Cody} \& {Sasselov}(2002)}]{cody02}
{Cody}, A.~M. \& {Sasselov}, D.~D. 2002, \apj, 569, 451

\bibitem[{{Deming} {et~al.}(2005{\natexlab{a}}){Deming}, {Brown},
  {Charbonneau}, {Harrington}, \& {Richardson}}]{deming05a}
{Deming}, D., {Brown}, T.~M., {Charbonneau}, D., {Harrington}, J., \&
  {Richardson}, L.~J. 2005{\natexlab{a}}, \apj, 622, 1149

\bibitem[{{Deming} {et~al.}(2005{\natexlab{b}}){Deming}, {Seager},
  {Richardson}, \& {Harrington}}]{deming05b}
{Deming}, D., {Seager}, S., {Richardson}, L.~J., \& {Harrington}, J.
  2005{\natexlab{b}}, \nat, 434, 740

\bibitem[{{Henry} {et~al.}(2000){Henry}, {Marcy}, {Butler}, \&
  {Vogt}}]{henry00}
{Henry}, G.~W., {Marcy}, G.~W., {Butler}, R.~P., \& {Vogt}, S.~S. 2000, \apjl,
  529, L41

\bibitem[{{Horne}(1986)}]{horne86}
{Horne}, K. 1986, \pasp, 98, 609

\bibitem[{{Kelsall} {et~al.}(1998){Kelsall}, {Weiland}, {Franz}, {Reach},
  {Arendt}, {Dwek}, {Freudenreich}, {Hauser}, {Moseley}, {Odegard},
  {Silverberg}, \& {Wright}}]{kelsall98}
{Kelsall}, T., {Weiland}, J.~L., {Franz}, B.~A., {Reach}, W.~T., {Arendt},
  R.~G., {Dwek}, E., {Freudenreich}, H.~T., {Hauser}, M.~G., {Moseley}, S.~H.,
  {Odegard}, N.~P., {Silverberg}, R.~F., \& {Wright}, E.~L. 1998, \apj, 508, 44

\bibitem[{{Knutson} {et~al.}(2006){Knutson}, {Charbonneau}, {Noyes}, {Brown},
  \& {Gilliland}}]{knutson06}
{Knutson}, H., {Charbonneau}, D., {Noyes}, R.~W., {Brown}, T.~M., \&
  {Gilliland}, R.~L. 2006, ArXiv Astrophysics e-prints

\bibitem[{{Krist}(1993)}]{krist93}
{Krist}, J. 1993, in ASP Conf. Ser. 52: Astronomical Data Analysis Software and
  Systems II, ed. R.~J. {Hanisch}, R.~J.~V. {Brissenden}, \& J.~{Barnes},
  536--+

\bibitem[{{Laughlin} {et~al.}(2005){Laughlin}, {Wolf}, {Vanmunster},
  {Bodenheimer}, {Fischer}, {Marcy}, {Butler}, \& {Vogt}}]{laughlin05}
{Laughlin}, G., {Wolf}, A., {Vanmunster}, T., {Bodenheimer}, P., {Fischer}, D.,
  {Marcy}, G., {Butler}, P., \& {Vogt}, S. 2005, \apj, 621, 1072

\bibitem[{{Mandel} \& {Agol}(2002)}]{mandel02}
{Mandel}, K. \& {Agol}, E. 2002, \apjl, 580, L171

\bibitem[{{Press} {et~al.}(1992){Press}, {Teukolsky}, {Vetterling}, \&
  {Flannery}}]{press92}
{Press}, W.~H., {Teukolsky}, S.~A., {Vetterling}, W.~T., \& {Flannery}, B.~P.
  1992, {Numerical Recipes in C: The Art of Scientific Computing}, 2nd edn.
  (Cambridge University Press)

\bibitem[{{Richardson} {et~al.}(2006){Richardson}, {Seager}, {Deming},
  {Harrington}, {Barry}, {Rajagopal}, \& {Danchi}}]{richardson05p2}
{Richardson}, L.~J., {Seager}, S., {Deming}, D., {Harrington}, J., {Barry},
  R.~K., {Rajagopal}, J., \& {Danchi}, W.~C. 2006, in IAUC 200, Direct
  Detection of Exoplanets: Science and Techniques, in press, ed. C.~{Aime} \&
  F.~{Vakili}

\bibitem[{{Rieke} {et~al.}(2004){Rieke}, {Young}, {Engelbracht}, {Kelly},
  {Low}, {Haller}, {Beeman}, {Gordon}, {Stansberry}, {Misselt}, {Cadien},
  {Morrison}, {Rivlis}, {Latter}, {Noriega-Crespo}, {Padgett}, {Stapelfeldt},
  {Hines}, {Egami}, {Muzerolle}, {Alonso-Herrero}, {Blaylock}, {Dole}, {Hinz},
  {Le Floc'h}, {Papovich}, {P{\'e}rez-Gonz{\'a}lez}, {Smith}, {Su}, {Bennett},
  {Frayer}, {Henderson}, {Lu}, {Masci}, {Pesenson}, {Rebull}, {Rho}, {Keene},
  {Stolovy}, {Wachter}, {Wheaton}, {Werner}, \& {Richards}}]{rieke04}
{Rieke}, G.~H., {Young}, E.~T., {Engelbracht}, C.~W., {Kelly}, D.~M., {Low},
  F.~J., {Haller}, E.~E., {Beeman}, J.~W., {Gordon}, K.~D., {Stansberry},
  J.~A., {Misselt}, K.~A., {Cadien}, J., {Morrison}, J.~E., {Rivlis}, G.,
  {Latter}, W.~B., {Noriega-Crespo}, A., {Padgett}, D.~L., {Stapelfeldt},
  K.~R., {Hines}, D.~C., {Egami}, E., {Muzerolle}, J., {Alonso-Herrero}, A.,
  {Blaylock}, M., {Dole}, H., {Hinz}, J.~L., {Le Floc'h}, E., {Papovich}, C.,
  {P{\'e}rez-Gonz{\'a}lez}, P.~G., {Smith}, P.~S., {Su}, K.~Y.~L., {Bennett},
  L., {Frayer}, D.~T., {Henderson}, D., {Lu}, N., {Masci}, F., {Pesenson}, M.,
  {Rebull}, L., {Rho}, J., {Keene}, J., {Stolovy}, S., {Wachter}, S.,
  {Wheaton}, W., {Werner}, M.~W., \& {Richards}, P.~L. 2004, \apjs, 154, 25

\bibitem[{{Seager} \& {Mall{\'e}n-Ornelas}(2003)}]{seager_mallen-ornelas03}
{Seager}, S. \& {Mall{\'e}n-Ornelas}, G. 2003, \apj, 585, 1038

\bibitem[{{Showman} \& {Guillot}(2002)}]{showman02}
{Showman}, A.~P. \& {Guillot}, T. 2002, \aap, 385, 166

\bibitem[{{Vernazza} {et~al.}(1976){Vernazza}, {Avrett}, \&
  {Loeser}}]{vernazza76}
{Vernazza}, J.~E., {Avrett}, E.~H., \& {Loeser}, R. 1976, \apjs, 30, 1

\bibitem[{{Vidal-Madjar} {et~al.}(2003){Vidal-Madjar}, {des Etangs}, {D{\'
  e}sert}, {Ballester}, {Ferlet}, {H{\' e}brard}, \& {Mayor}}]{vidal-madjar03}
{Vidal-Madjar}, A., {des Etangs}, A.~L., {D{\' e}sert}, J.-M., {Ballester},
  G.~E., {Ferlet}, R., {H{\' e}brard}, G., \& {Mayor}, M. 2003, \nat, 422, 143

\bibitem[{{Werner} {et~al.}(2004){Werner}, {Roellig}, {Low}, {Rieke}, {Rieke},
  {Hoffmann}, {Young}, {Houck}, {Brandl}, {Fazio}, {Hora}, {Gehrz}, {Helou},
  {Soifer}, {Stauffer}, {Keene}, {Eisenhardt}, {Gallagher}, {Gautier}, {Irace},
  {Lawrence}, {Simmons}, {Van Cleve}, {Jura}, {Wright}, \&
  {Cruikshank}}]{werner04}
{Werner}, M.~W., {Roellig}, T.~L., {Low}, F.~J., {Rieke}, G.~H., {Rieke}, M.,
  {Hoffmann}, W.~F., {Young}, E., {Houck}, J.~R., {Brandl}, B., {Fazio}, G.~G.,
  {Hora}, J.~L., {Gehrz}, R.~D., {Helou}, G., {Soifer}, B.~T., {Stauffer}, J.,
  {Keene}, J., {Eisenhardt}, P., {Gallagher}, D., {Gautier}, T.~N., {Irace},
  W., {Lawrence}, C.~R., {Simmons}, L., {Van Cleve}, J.~E., {Jura}, M.,
  {Wright}, E.~L., \& {Cruikshank}, D.~P. 2004, \apjs, 154, 1

\bibitem[{{Winn} \& {Holman}(2005)}]{winn_holman05}
{Winn}, J.~N. \& {Holman}, M.~J. 2005, \apjl, 628, L159

\bibitem[{{Wittenmyer} {et~al.}(2005){Wittenmyer}, {Welsh}, {Orosz}, {Schultz},
  {Kinzel}, {Kochte}, {Bruhweiler}, {Bennum}, {Henry}, {Marcy}, {Fischer},
  {Butler}, \& {Vogt}}]{wittenmyer05}
{Wittenmyer}, R.~A., {Welsh}, W.~F., {Orosz}, J.~A., {Schultz}, A.~B.,
  {Kinzel}, W., {Kochte}, M., {Bruhweiler}, F., {Bennum}, D., {Henry}, G.~W.,
  {Marcy}, G.~W., {Fischer}, D.~A., {Butler}, R.~P., \& {Vogt}, S.~S. 2005,
  \apj, 632, 1157

\end{thebibliography}

%Table 1: photometry
\clearpage

\begin{deluxetable}{cccc}
\tablecaption{Calibrated, unbinned photometry.
\label{tbl:phot}}
\tablewidth{0pt}
\tablehead{
\colhead{HJD} & \colhead{Phase} & \colhead{Relative Intensity} & \colhead{Error}}
\startdata
%2453344.6466211 & -0.04063 & 1.009855 & 0.008501 \\
%2453344.6467485 & -0.04060 & 0.989528 & 0.008168 \\
%2453344.6468759 & -0.04056 & 1.001790 & 0.008869 \\
%2453344.6470034 & -0.04052 & 1.003001 & 0.008082 \\
%2453344.6471308 & -0.04049 & 0.978126 & 0.008656 \\
2453344.6466211 & -0.03728 & 1.009530 & 0.008265 \\
2453344.6467485 & -0.03724 & 1.001394 & 0.008643 \\
2453344.6468759 & -0.03721 & 1.010476 & 0.008172 \\
2453344.6470034 & -0.03717 & 1.012307 & 0.008398 \\
2453344.6471308 & -0.03713 & 0.992136 & 0.008321 \\
\enddata

[The complete version of this table is in the electronic edition of
the Journal.  The printed edition contains only a sample.]

\end{deluxetable}

%Table 2: best fit from straight-line segments
\clearpage

\begin{deluxetable}{lccc}
\tablecaption{Observables derived from $\chi_\nu^2$ minimization of simple eclipse curves.
\label{tbl:obs}}
\tablewidth{0pt}
\tablehead{
\colhead{Parameter} & \colhead{Method 1} & \colhead{Method 2} & \colhead{Error} } 
\startdata
Depth & 0.01496 & 0.01493 & 0.00029 \\
$t_{\mbox{\scriptsize T}}$ (hr) & 2.978 & 2.979 & 0.051 \\
$t_{\mbox{\scriptsize F}}$ (hr) & 2.254 & 2.253 & 0.058 \\
Time Offset (hr) & 0.000 & 0.001 & 0.013 \\
$\chi_\nu^2$ & 1.00514 & 1.00598 & - \\
\enddata
\end{deluxetable}

%Table 3: derived planetary parameters and errors
\clearpage

\begin{deluxetable}{llll}
%\tablecaption{Derived physical parameters from best fit to straight-line segments, for three assumed values of the stellar mass.  These parameters were derived for a range of stellar masses from 0.93 to 1.30~M$_\sun$. (See Figure~\ref{fig:mstar}.)
\tablecaption{Derived physical parameters for the two minimization techniques.
\label{tbl:params} }
\tablewidth{0pt}
\tablehead{
\colhead{Parameter} & \colhead{Method 1} & \colhead{Method 2} & \colhead{Error}}
\startdata
%Assumed Stellar Mass (M$_\sun$) & 0.93 & 1.10 & 1.19 & - \\
%Stellar Radius (R$_\sun$) & 1.031 & 1.090 & 1.119 & 0.066 \\
%Planetary Radius (R$_J$) & 1.229 & 1.300 & 1.334 & 0.083 \\
%Orbital Inclination (deg) & 87.24 & 87.24 & 87.24 & 0.76 \\
%Semi-Major Axis (AU) & 0.0443 & 0.0468 & 0.0480 & 0. \\
Assumed Stellar Mass (M$_\sun$) & 1.173 & 1.171 & - \\
Stellar Radius (R$_\sun$) & 1.063 & 1.064 & 0.069 \\
Planetary Radius (R$_{\mbox{\scriptsize J}}$) & 1.265 & 1.265 & 0.085 \\
Orbital Inclination (deg) & 88.00 & 87.97 & 0.85 \\
%Semi-Major Axis (AU) & & & \\
\enddata
\end{deluxetable}

%Table 4: observed time of center eclipse
\clearpage

\begin{deluxetable}{lcc}
\tablecaption{Observed time of center eclipse.
\label{tbl:ecltime} }
\tablewidth{0pt}
\tablehead{
\colhead{} & \colhead{Time (HJD)} & \colhead{Error} }
\startdata
Event~1 (Half) & 2453344.768245 & 0.002608\\
Event~2 (Full) & 2453549.201422 & 0.000617\\
\enddata
\end{deluxetable}

%% Text for table notes should follow after the \enddata but before
%% the \end{deluxetable}. Make sure there is at least one \tablenotemark
%% in the table for each \tablenotetext.

%\tablecomments{Table \ref{tbl-1} is published in its entirety in the electronic edition of the {\it Astrophysical Journal}. A portion is shown here for guidance regarding its form and content.}

%\tablenotetext{a}{Sample footnote for table~\ref{tbl-1} that was generated with the deluxetable environment}
%\tablenotetext{b}{Another sample footnote for table~\ref{tbl-1}}

\clearpage

%%% Figure 1--show the data

\clearpage

%%% Figure 2--show the contour plots

\clearpage

%%% Figure 3--show the radii vs. assumed stellar mass

\end{document}